%% file: main.tex
\pdfoutput=0
\documentclass[conference]{IEEEtran}

\ifCLASSOPTIONonecolumn
\usepackage{setspace}
\fi

\IEEEoverridecommandlockouts
\usepackage{cite}
\usepackage{amsmath,amssymb,amsfonts}
\usepackage{adjustbox}
\usepackage{graphicx}
\usepackage{booktabs}
\usepackage{multirow}
\usepackage{color}
\usepackage{comment}
\usepackage{psfig}
\usepackage{multirow}
\usepackage{placeins}
\newcommand{\mycomment}[1]{%
}%
\usepackage{environ}
\NewEnviron{nestedcomment}{\mycomment{\BODY}}
\usepackage{algorithm}
\usepackage{algpseudocode}
\ifCLASSOPTIONonecolumn
\usepackage[font=small,labelfont=bf]{caption}
\usepackage{etoolbox}
\makeatletter
\patchcmd{\@makecaption}
  {\scshape}
  {}
  {}
  {}
\makeatother

\fi
\def\urltilde{\kern -.15em\lower .7ex\hbox{\~{}}\kern .04em}

\begin{document}
\title{Alternative Channel Charting Techniques in Cellular Wireless Communications
\thanks{This work is partially supported by NSF grant 2030029.}
}

\ifCLASSOPTIONonecolumn
\author{\IEEEauthorblockN{ Yonghong Jiang}\\
\IEEEauthorblockA{\textrm{CPCC, Department of EECS} \\
\textrm{University of California, Irvine, CA, USA}\\
alyas@uci.edu\\[7mm]}
\and
\IEEEauthorblockN{ Ender Ayanoglu, {\em Fellow, IEEE}}\\
\IEEEauthorblockA{\textrm{CPCC, Department of EECS} \\
\textrm{University of California, Irvine, CA, USA}
 \\ayanoglu@uci.edu
}}
\else
\author{\IEEEauthorblockN{ Yonghong Jiang}
\IEEEauthorblockA{\textrm{CPCC, Department of EECS} \\
\textrm{University of California, Irvine, CA, USA}\\
alyas@uci.edu\\}

\and
\IEEEauthorblockN{Ender Ayanoglu, {\em Fellow, IEEE}}
\IEEEauthorblockA{\textrm{CPCC, Department of EECS} \\
\textrm{University of California, Irvine, CA, USA}\\
ayanoglu@uci.edu
}
}
\fi

\maketitle

\begin{abstract}
We investigate the use of conventional angle of arrival (AoA) algorithms the Bartlett's algorithm, the Minimum Variance Distortion Response (MVDR or Capon) algorithm, and the Minimum Norm algorithm for estimating the AoA $\theta$ together with our previously introduced algorithms linear regression (LR), inverse of the root sum squares of channel coefficients (ISQ), as well as a novel use of the MUSIC algorithm for estimating the distance from the base station, $\rho$ in the context of channel charting. We carry out evaluations in terms of the visual quality of the channel charts, the dimensionality reduction performance measures trustworthiness (TW) and connectivity (CT), as well as the execution time of the algorithms. We find that although the Bartlett's algorithm, MVDR, and Minimum Norm algorithms have sufficiently close performance to techniques we studied earlier, the Minimum Norm algorithm has significantly higher computational complexity than the other two. Previously, we found that the use of the MUSIC algorithm for estimation of both $\theta$ and $\rho$ has a very high performance. In this paper, we investigated and quantified the performance of the Bartlett algorithm in its use for estimating both $\theta$ and $\rho$, similar to the our previously introduced technique of using MUSIC for estimating both.
\end{abstract}

\input{introduction}

\input{channelmodels}
\input{estimation}

\input{environment}

\input{performance}

\input{AoAAlgorithms}
\input{simulationresults}
\input{conclusion}

\ifCLASSOPTIONonecolumn
\clearpage\newpage
\fi
\bibliographystyle{IEEEtran}
\bibliography{ref}
\end{document}

%% file: introduction.tex
\section{Introduction}\label{ch:1}
A channel chart is a chart created from channel state information (CSI). It has the property of preserving the relative geometry of the radio environment consisting of a base station (BS) and user equipments (UEs) \cite{b1}. By employing this chart, the BS locates the relative locations of the UEs. This has the potential of enabling many applications such as handover, cell search, user localization, etc.
While most of the works on this subject employed estimation of a channel chart using dimensionality reduction techniques, in this paper we calculate the channel chart directly by using model-based approaches.

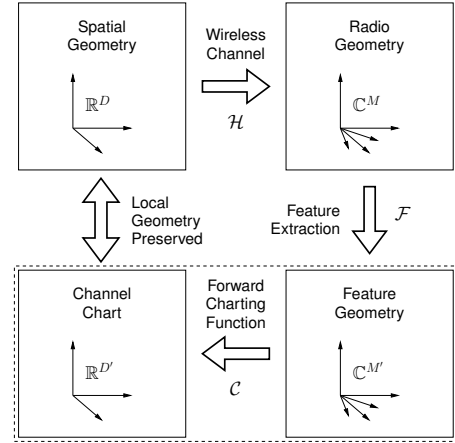
\begin{figure}[!ht]%[htbp]
\centering
\ifCLASSOPTIONonecolumn
\scalebox{0.5}{\input{cchart.pstex_t}}
\else
\scalebox{0.35}{\input{cchart.pstex_t}}
\fi
\caption{Summary of channel charting via dimensionality reduction \cite{b1}.}
\label{fig:cchart}
\end{figure}
We will begin our discussion by Fig.~\ref{fig:cchart}, which is a redrawn and simplified version of \cite[Fig.~3]{b1}. UE transmitters are located in spatial geometry $\mathbb{R}^D$, where $D=2$ or $3$ \cite{b1}. The BS receiver calculates CSI in radio geometry $\mathbb{C}^M$ where $M \gg D$.
Then, a channel chart is created in $\mathbb{R}^{D'}$ where $D'\le D$ such that
the representation in $\mathbb{R}^{D'}$ preserves the local geometry of the original spatial locations in $\mathbb{R}^D$, in other words,
the relative positions of the UEs. Reference \cite{b1} introduces and compares three dimensionality reduction algorithms, namely principal component analysis (PCA), Sammon's mapping (SM), and autoencoder (AE). PCA is a linear technique for dimensionality reduction. It maps a high-dimensional point set (e.g., CSI features) into a low-dimensional point set (e.g., the channel chart) in an unsupervised approach. It does so by performing dimensionality reduction only for the data points used in calculations. It does not form a function one can use to perform dimensionality reduction for future data points. For that reason, strictly speaking, it is not a machine learning (ML) algorithm, although sometimes it is quoted as an unsupervised ML algorithm. SM is a nonlinear method for dimensionality reduction which retains small pairwise distances between the two point sets \cite{b1}. Similarly to PCA, it does not form a function for dimensionality reduction of future data points. Whereas, an AE is a deep artificial neural network used for unsupervised dimensionality reduction \cite{b1}. Unlike PCA and SM, it does perform learning. Thus, it can be used for future data points.

Consider Fig.~\ref{fig:cchart}. In this figure, there are four blocks to carry out channel charting. In the upper left, the spatial geometry in $\mathbb{R}^D$ is depicted. In the upper right block, the radio geometry in $\mathbb{C}^M$ is calcaulated. The lower blocks perform feature extraction and forward charting to create channel charts. In the approach in this paper we keep the upper two blocks, as in \cite{AA24,AA23a,AA23}. In our approach, one replaces the lower two blocks with model-based techniques to directly determine the angle of arrival and the distance from the BS of the UE. Thus, the relative positions of the UEs with respect to the BS are preserved, and the basic goal of channel charting is automatically satisfied.

%% file: cchart.pstex_t
\begin{picture}(0,0)%
\includegraphics{cchart.pstex}%
\end{picture}%
\setlength{\unitlength}{3947sp}%
\begingroup\makeatletter\ifx\SetFigFont\undefined%
\gdef\SetFigFont#1#2#3#4#5{%
  \reset@font\fontsize{#1}{#2pt}%
  \fontfamily{#3}\fontseries{#4}\fontshape{#5}%
  \selectfont}%
\fi\endgroup%
\begin{picture}(7974,7899)(814,-8248)
\put(2401,-886){\makebox(0,0)[b]{\smash{{\SetFigFont{17}{20.4}{\sfdefault}{\mddefault}{\updefault}{\color[rgb]{0,0,0}Spatial}%
}}}}
\put(2401,-1216){\makebox(0,0)[b]{\smash{{\SetFigFont{17}{20.4}{\sfdefault}{\mddefault}{\updefault}{\color[rgb]{0,0,0}Geometry}%
}}}}
\put(7201,-886){\makebox(0,0)[b]{\smash{{\SetFigFont{17}{20.4}{\sfdefault}{\mddefault}{\updefault}{\color[rgb]{0,0,0}Radio}%
}}}}
\put(7201,-1216){\makebox(0,0)[b]{\smash{{\SetFigFont{17}{20.4}{\sfdefault}{\mddefault}{\updefault}{\color[rgb]{0,0,0}Geometry}%
}}}}
\put(2401,-5686){\makebox(0,0)[b]{\smash{{\SetFigFont{17}{20.4}{\sfdefault}{\mddefault}{\updefault}{\color[rgb]{0,0,0}Channel}%
}}}}
\put(2401,-6016){\makebox(0,0)[b]{\smash{{\SetFigFont{17}{20.4}{\sfdefault}{\mddefault}{\updefault}{\color[rgb]{0,0,0}Chart}%
}}}}
\put(4801,-1036){\makebox(0,0)[b]{\smash{{\SetFigFont{17}{20.4}{\sfdefault}{\mddefault}{\updefault}{\color[rgb]{0,0,0}Wireless}%
}}}}
\put(4801,-1366){\makebox(0,0)[b]{\smash{{\SetFigFont{17}{20.4}{\sfdefault}{\mddefault}{\updefault}{\color[rgb]{0,0,0}Channel}%
}}}}
\put(4801,-5536){\makebox(0,0)[b]{\smash{{\SetFigFont{17}{20.4}{\sfdefault}{\mddefault}{\updefault}{\color[rgb]{0,0,0}Forward}%
}}}}
\put(4801,-5866){\makebox(0,0)[b]{\smash{{\SetFigFont{17}{20.4}{\sfdefault}{\mddefault}{\updefault}{\color[rgb]{0,0,0}Charting}%
}}}}
\put(4801,-6196){\makebox(0,0)[b]{\smash{{\SetFigFont{17}{20.4}{\sfdefault}{\mddefault}{\updefault}{\color[rgb]{0,0,0}Function}%
}}}}
\put(7201,-6016){\makebox(0,0)[b]{\smash{{\SetFigFont{17}{20.4}{\sfdefault}{\mddefault}{\updefault}{\color[rgb]{0,0,0}Geometry}%
}}}}
\put(7201,-5686){\makebox(0,0)[b]{\smash{{\SetFigFont{17}{20.4}{\sfdefault}{\mddefault}{\updefault}{\color[rgb]{0,0,0}Feature}%
}}}}
\put(2926,-4036){\makebox(0,0)[lb]{\smash{{\SetFigFont{17}{20.4}{\sfdefault}{\mddefault}{\updefault}{\color[rgb]{0,0,0}Local}%
}}}}
\put(2926,-4366){\makebox(0,0)[lb]{\smash{{\SetFigFont{17}{20.4}{\sfdefault}{\mddefault}{\updefault}{\color[rgb]{0,0,0}Geometry}%
}}}}
\put(2926,-4696){\makebox(0,0)[lb]{\smash{{\SetFigFont{17}{20.4}{\sfdefault}{\mddefault}{\updefault}{\color[rgb]{0,0,0}Preserved}%
}}}}
\put(6676,-4186){\makebox(0,0)[rb]{\smash{{\SetFigFont{17}{20.4}{\sfdefault}{\mddefault}{\updefault}{\color[rgb]{0,0,0}Feature}%
}}}}
\put(6676,-4516){\makebox(0,0)[rb]{\smash{{\SetFigFont{17}{20.4}{\sfdefault}{\mddefault}{\updefault}{\color[rgb]{0,0,0}Extraction}%
}}}}
\put(7651,-4261){\makebox(0,0)[lb]{\smash{{\SetFigFont{20}{24.0}{\familydefault}{\mddefault}{\updefault}{\color[rgb]{0,0,0}${\cal F}$}%
}}}}
\put(4801,-2611){\makebox(0,0)[b]{\smash{{\SetFigFont{20}{24.0}{\familydefault}{\mddefault}{\updefault}{\color[rgb]{0,0,0}${\cal H}$}%
}}}}
\put(4801,-7411){\makebox(0,0)[b]{\smash{{\SetFigFont{20}{24.0}{\familydefault}{\mddefault}{\updefault}{\color[rgb]{0,0,0}${\cal C}$}%
}}}}
\put(2101,-2311){\makebox(0,0)[lb]{\smash{{\SetFigFont{20}{24.0}{\familydefault}{\mddefault}{\updefault}{\color[rgb]{0,0,0}$\mathbb{R}^D$}%
}}}}
\put(2101,-7111){\makebox(0,0)[lb]{\smash{{\SetFigFont{20}{24.0}{\familydefault}{\mddefault}{\updefault}{\color[rgb]{0,0,0}$\mathbb{R}^{D'}$}%
}}}}
\put(6901,-2311){\makebox(0,0)[lb]{\smash{{\SetFigFont{20}{24.0}{\familydefault}{\mddefault}{\updefault}{\color[rgb]{0,0,0}$\mathbb{C}^M$}%
}}}}
\put(6901,-7111){\makebox(0,0)[lb]{\smash{{\SetFigFont{20}{24.0}{\familydefault}{\mddefault}{\updefault}{\color[rgb]{0,0,0}$\mathbb{C}^{M'}$}%
}}}}
\end{picture}%

%% file: channelmodels.tex
\section{Channel Models}
We employ three channel models, namely vanilla line-of-sight (LOS), Quadriga LOS (QLOS), and Quadriga  non-LOS (QNLOS) \cite{b4,b44,Jaeckel16}. These are the same models used in \cite{b1}, as well as our papers \cite{AA24,AA23a,AA23}. We start with the simplest, vanilla LOS. Vanilla LOS is one LOS ray
described as
\begin{equation}
h=\rho^{-r}\; e^{-j\left(\frac{2 \pi \rho}{\lambda}+\phi\right)} %+ n
\label{eq20}
\end{equation}
where $\rho$ is the distance between the transmitter and the receiver and $r$ is known as the path loss exponent. %, and $n \sim {\cal CN}(0,\sigma_n^2)$.
%\rev{As can be seen from (\ref{eq20}), the magnitude and phase of the channel are deterministic, and can be determined by distance only.}
In (\ref{eq20}), the first term in the channel phase is linearly proportional with the distance $\rho$. % between the transmitter and the receiver.
The second term $\phi$ is a uniformly distributed random variable in $[0, 2\pi)$. The channel amplitude is a random variable (Rician (QLOS) or Rayleigh (QNLOS)) which is inversely proportional to the distance square for free space, ${\sim}\rho^{-2}$, i.e., the path loss exponent $r = 2$.

\ifCLASSOPTIONonecolumn
\begin{table}[!t]
\begin{center}
\begin{tabular}{||c | c||}
 \hline
 Parameter & Value \\ [0.5ex]
 \hline\hline
 Antenna array & Uniform Linear Array (ULA) with spacing  $\lambda/2 = 7.495$ cm \\
 \hline
 Number of array antennas & 32  \\
 \hline
 Number of transmitters (UEs)& 2048 \\
 \hline
 Carrier frequency & 2.0 GHz \\
 \hline
Bandwidth & 312.5 kHz \\
 \hline
 Number of clusters &0  \\ %[1ex]
 \hline
  Number of subcarriers &1 (up to 32 in the case of the MM algorithm (Sec.~\ref{sec:MM}))  \\ %[1ex]
 \hline
\end{tabular}
\end{center}
\caption{Simulation parameters.}\label{tab1}
\end{table}
\else
\begin{table}[!t]
\begin{center}
\caption{Simulation parameters.}\label{tab1}
\begin{tabular}{||c | c||}
 \hline
 Parameter & Value \\ [0.5ex]
 \hline\hline
 Antenna array & Uniform Linear Array (ULA)\\
               & with spacing  $\lambda/2 = 7.495$ cm \\
 \hline
 Number of array antennas & 32  \\
 \hline
 Number of transmitters (UEs)& 2048 \\
 \hline
 Carrier frequency & 2.0 GHz \\
 \hline
Bandwidth & 312.5 kHz \\
 \hline
Number of clusters &0  \\ %[1ex]
 \hline
Number of subcarriers &1 (up to 32 in the case of the \\ %[1ex]
&MM algorithm (Sec.~\ref{sec:MM}))  \\ %[1ex]
 \hline
\end{tabular}
\end{center}
\end{table}
\fi
Next we discuss the Quadriga channel model \cite{b4,b44,Jaeckel16}. Quadriga stands for quasi deterministic radio channel generator. It is a statistical three-dimensional geometry-based stochastic channel model employing ray tracing. According to \cite{b4}, it has the following features: {\em i)\/} three-dimensional propagation (antenna modeling, geometric polarization, scattering clusters), {\em ii)\/} continuous-time evolution, {\em iii)\/} spatially correlated large- and small-scale fading, and {\em iv)\/} transition between varying propagation scenarios. The Quadriga model is very customizable. It has many features and details. %Because of its very detailed nature, we refer the reader to \cite{b4,b44,Jaeckel16}.
The model was validated by measurements in downtown Dresden, Germany \cite[Ch. 4]{Jaeckel16} and in downtown Berlin, Germany \cite[Ch. 5]{Jaeckel16}.
In this paper we used the parameters in Table~\ref{tab1} with the Urban Macro-Cell (UMa) version of the Quadriga mode in the simulations. Some details of the measurement setup are available in \cite[Sec.~III]{b4}, in specific detail in \cite[Table~II]{b4}.

The signal-to-noise ratio (SNR) in channel model is calculated by considering the power in the received signal ($P_r$) and the power in the noise measured at the receiver ($P_n$). We note that while the estimated channel would have some noise added to it, the most significant component of the noise at the receiver is additive white Gaussian thermal noise. Then, the SNR at the receiver is given as $\textrm{SNR} = P_r/P_n$ where $P_r$ takes into account the transmitted power and the channel model, see, e.g., Sec.~II-B in \cite{MYPC22}. In the code \cite{ChaChaCode} which we used as the basis for our simulations, the calculation of SNR is carried out by normalizing $P_r$ and then properly scaling the additive white Gaussian thermal noise power $P_n$ for all three channel models.

%% file: estimation.tex
\section{Estimating the Coordinates ${\theta}$ and ${\rho}$}\label{ch:2}
We will use the symbol $\theta$ for the angle of arrival (AOA) and $\rho$ for the distance between the BS and the UE. Note that one can estimate $\theta$ and $\rho$ concurrently because they do not depend on each other. In this section, we will first discuss how to estimate $\theta$ by using the MUSIC algorithm and then we will discuss three algorithms to estimate $\rho$.
\subsection{Estimating $\theta$ Using MUSIC}
\begin{figure}[!tb]
\vspace{18mm}
\centering
\ifCLASSOPTIONonecolumn
\includegraphics[bb = 0 0 612 792, width=0.185\textwidth]{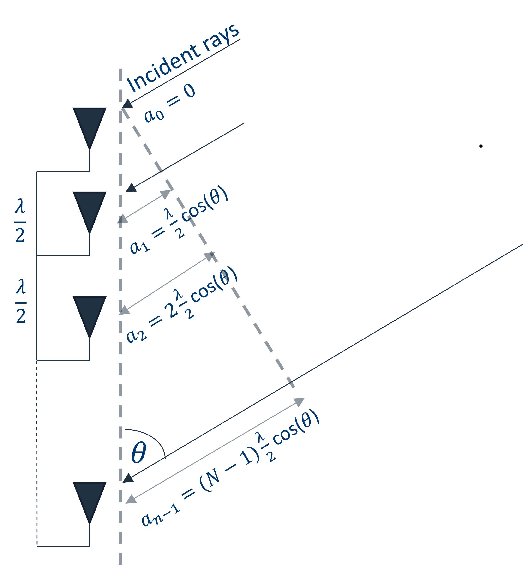}
\else
\vspace{-1.8cm}
\includegraphics[bb = 0 0 200 273, clip, width=0.25\textwidth]{fig1.eps}
% -0 -0 251 273
\fi
\caption{Angle of arrival $(\theta)$ relation with phase.}
\label{fig1}
\end{figure}
Consider Fig.~\ref{fig1}. One can see from this figure that each antenna element receives a ray which travels an additional distance $\frac{\lambda}{2} \cos(\theta)$ as compared to the previous element. As a result, the incremental phase shift for each antenna element is $e^{j\pi \cos(\theta)}$. Thus, one can compose the steering vector
\begin{equation}
{\bf A}(\theta)= (1,e^{j\pi\cos(\theta)},e^{j\pi 2\cos(\theta)},\ldots,e^{j\pi (N_R-1)\cos(\theta)})^T,
\label{eq230}
\end{equation}
where $N_R$ is the number of receive antennas at the BS. This vector is employed in determining the AOA as well as in beamforming applications.
\begin{algorithm}[!t]
\caption{MUSIC Procedure for Estimating $\theta$}\label{alg:cap}
\begin{algorithmic}
    \State Calculate the CSI across antennas and subcarriers covariance matrix ${\bf R}$ = $\mathbb{E}[{\bf h}{\bf h}^{H}]$
    \State  Get the eigenvectors and eigenvalues of {\bf R}
    \State Separate system subspace $\cal S$ and noise subspace $\cal N$ by defining a threshold
    \State Calculate ${\bf N}$ by concatenating the eigenvectors of $\cal N$
         \For{\texttt{$\theta = 0: 180$} in increments of 1}
            \State Calculate the steering vector ${\bf A}(\theta)$
        \State Calculate the $\textrm{PMF}(\theta)=\frac{1}{\textrm{Norm}_2({\bf N}^H{\bf A}(\theta))}$
      \EndFor
    \State Search the PMF for a peak and find the corresponding $\theta$
\end{algorithmic}
\end{algorithm}
The steering vector ${\bf A}(\theta)$ is embedded within the CSI correlation matrix (${\bf R} =\mathbb{E} [{\bf h}{\bf h}^H]$), where ${\bf h}$ is the received channel vector at the BS along with noise. The vector ${\bf h}$ is $N_R\times 1$ where $N_R$ is the number of antennas at the BS.
By decomposing $\bf R$ into its eigenvectors and examining the corresponding eigenvalues, the eigenvectors can be separated into a signal subspace $\cal S$ and a noise subspace $\cal N$. This is achieved by using the fact that the noise eigenvectors will correspond to very small eigenvalues compared to the signal space eigenvalues. The two subspaces $\cal S$ and $\cal N$ are orthogonal. Let's assume that the dimensionality of the noise subspace $\cal N$ is $p$. Form the matrix ${\bf N}$ (dimension $N_R \times p$) by concatenating the eigenvectors of $\cal N$. The multiplication of the noise subspace eigenvectors matrix $\bf N$ and the steering vector will be almost zero. We can use this concept to find the correct angle by sweeping $\theta$ in the steering vector as described in Algorithm~\ref{alg:cap} where PMF($\theta$) is a probability mass function within a scale of constant.
%We provide a flowchart of the algorithm on left hand side in Figure~\ref{fig25}.
%
\subsection{Estimating $\rho$}
We will now discuss three methods on how to estimate $\rho$ \cite{AA24}. Please refer to (\ref{eq20}) with $r=2$. This is the simple channel ray model we will employ below.
\subsubsection{Estimating $\rho$ Using ISQ}
In this method, we calculate the square root inverse of the sum of CSI magnitudes for all antennas as
\begin{equation}
\rho=\frac{1}{\sqrt{\sum_{n=0}^{N_R-1} {\rm abs}(h_n)}} , \label{eq3}
\end{equation}
where $h_n$ is the channel between the UE and the $n$-th antenna at the BS and $N_R$ is the number of antennas at the base station.
We call this algorithm as ISQ (inverse square root sum). The algorithm is motivated by (\ref{eq20}) with the path loss component $r=2$. In (\ref{eq3}), $\rho = {1}/\sqrt{{\rm abs}(h_n)}$ is calculating the average $\rho$.\footnote{Note that
\begin{equation}
\rho' =\frac{1}{\sqrt{\frac{1}{N_R}\sum_{n=0}^{N_R-1} {\rm abs}(h_n)}} = \sqrt{N_R} \rho.
\end{equation}
In other words, the true average $\rho'$ is proportional to $\rho$. Therefore, estimated $\rho$ is not to scale with the real $\rho$, but that will not affect the TW and CT.} % as we will see later.}
\subsubsection{Estimating $\rho$ Using LR}
%
\begin{comment}
\begin{figure*}[!t]
\centering
\resizebox{\textwidth}{!}{
  \renewcommand{\arraystretch}{0}%
  \begin{tabular}{@{}c@{\hspace{0.25pt}}c@{}}
  \includegraphics[width=0.25in]{unsupervised_scatter.eps} &
  \includegraphics[width=0.25in]{regression_scatter.eps} \\
  \end{tabular}
}
\caption{Correlation of real $\rho$ vs estimated $\rho$ under the channel model QNLOS.}
\label{fig20}
\end{figure*}
\begin{comment}
\end{comment}
This is a learning-based, supervised approach. It is assumed that the location of 256 (out of 2048) UEs are known. Then, a linear regression is carried out with the logarithm of the sum of CSI magnitudes for all antennas to find $a$ and $b$ in
\begin{equation}
\rho=aX+b,\ \ \text{where}\ X=\log\sum_{n=0}^{N_R-1} {\rm abs}(h_n) . \label{eq4}
\end{equation}
For the first 256 UEs, we carry out a linear regression and use the known $\rho$ and $X$ values to generate $a$ and $b$. Then for the rest of the UEs, we use (\ref{eq4}) to estimate $\rho$ based on their $X$ values.
We name this algorithm the LR algorithm. The unsupervised performance of the ISQ algorithm is almost identical to the LR algorithm \cite{AA24}. Noting the $\log$ operation in (\ref{eq4}), and the fact that linear regression will generate $a<0$, this is a different way of expressing (\ref{eq3}).
\footnote{Note that
\begin{equation}
X'=\log\bigg(\frac{1}{N_R}\sum_{n=0}^{N_R-1} {\rm abs}(h_n)\bigg) = X - \log(N_R).\nonumber
\end{equation}
Therefore, the true average $X'$ differs from $X$ by a constant term, which can be absorbed by $b$ in (\ref{eq4}).} %We would like to point out to the subtlety that while the LR algorithm employs the model in (\ref{eq4}), because of the use of linear regression being based on the first 256 (in our case) UE locations, it may be considered training based. We also note that the regression in (\ref{eq4}) is called a linear-log regression. Another approach could be to use log-log regression where $\rho$ in the left hand side of (\ref{eq4}) is replaced by $\log\rho$ \cite{Benoit11}. Although this is closer to the model in (\ref{eq3}), we did not observe a significant difference in numerical simulations considering linear-log and log-log regression techniques in terms of CT and TW. %This may not be surprising since the log curve can be considered close to linear in its pre-saturated region.
%In addition, as can be observed from Fig.~\ref{fig20}, both ISQ and LR generate estimates that correlate linearly with real $\rho$.}
%
\subsubsection{Estimating $\rho$ Using MUSIC}\label{sec:MM}
\begin{figure}[!t]%[htbp]
  \centering
  \includegraphics[bb = 0 0 651 572, width=0.4\textwidth]{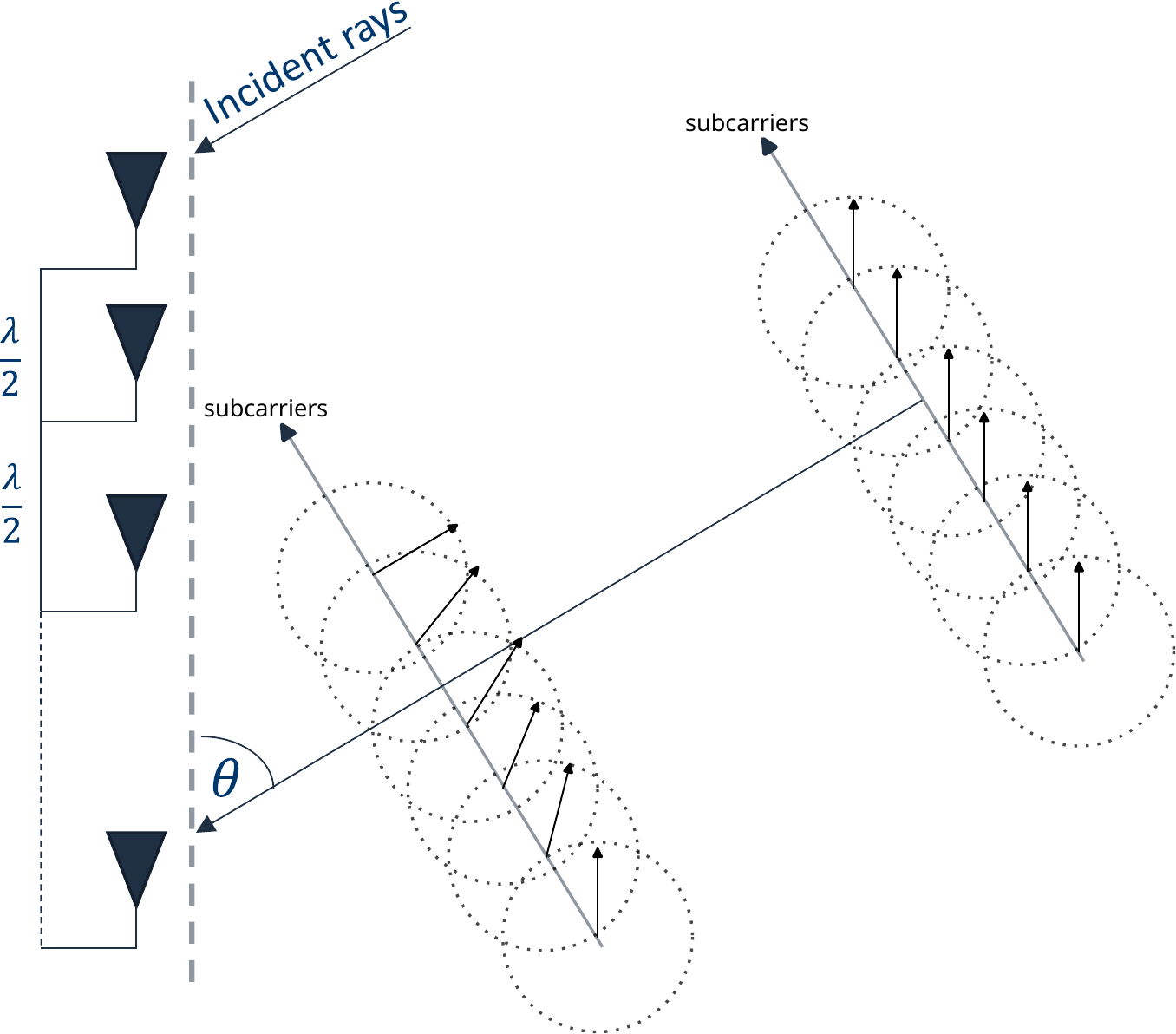}
  \caption{Phase change across subcarriers with distance.}
  \label{fig24}
\end{figure}
\begin{algorithm}[!t]
\caption{MUSIC Procedure for Estimating $\rho$}\label{alg:cap2}
\begin{algorithmic}
    \State Calculate the CSI across antennas and subcarriers covariance matrix ${\bf R}$ = $\mathbb{E}[{\bf h}{\bf h}^{H}]$
    \State  Get the eigenvectors and eigenvalues of {\bf R}
    \State Separate system subspace $\cal S$ and noise subspace $\cal N$ by defining a threshold
    \State Calculate ${\bf N}$ by concatenating the eigenvectors of $\cal N$
         \For{\texttt{$\rho = 0: 1000$} in increments of 1}
            \State Calculate vector ${\bf B}(\rho)$
        \State Calculate the $\textrm{PMF}(\rho)=\frac{1}{\textrm{Norm}_2({\bf N}^H {\bf B}(\rho))}$
      \EndFor
    \State Search the PMF for a peak and find the corresponding $\rho$
\end{algorithmic}
\end{algorithm}
For this algorithm, we use the same principle for estimating $\rho$ as in estimating $\theta$. We assume the transmission is multicarrier-based, and we use MUSIC to make use the phase difference among subcarriers. Please refer to Fig.~\ref{fig24}. Note that as the ray travels, the phases of the subcarriers change with rate according to their frequencies. If the subcarriers have a spacing of $\Delta\! f$ and we have $N_S$ subcarriers, their phase relation with distance is given as
\begin{equation}
{\bf B}(\rho)= (1,e^{-j2\pi\rho\Delta\!{f}/c},e^{-j2\pi\rho 2\Delta\!{f}/c},\ldots,e^{-j2\pi\rho (N_s-1) \Delta\!{f}/c})^T
\label{eq23}
\end{equation}
where %$\Delta{f}$ is the spacing of the subcarriers,
$\rho$ is the distance %$N_s$ is the number of subcarriers,
and $c$ is the speed of light.
The vector ${\bf B}(\rho)$ will be used exactly as we used the steering vector ${\bf A}(\theta)$ in estimating $\theta$. The procedure is explained in Algorithm~\ref{alg:cap2}. We call the combination of using MUSIC to estimate $\theta$ and using MUSIC to estimate $\rho$ the MUSIC/MUSIC (MM) algorithm. %A flowchart for this algorithm is given in Fig.~\ref{fig25}. In this flowchart, $N_{NS}$ is the number of noise subspace eigenvectors, i.e., $N_{NS}$ eigenvectors are selected for noise subspace.

%% file: environment.tex
\section{Simulation Environment and Basis for Comparison}\label{ch:3}
\ifCLASSOPTIONonecolumn
\begin{figure*}[!t]
\centering
  \renewcommand{\arraystretch}{0}%
  \begin{tabular}{@{}c@{\hspace{1pt}}c@{\hspace{1pt}}c@{}}
  \includegraphics[width=3.0in]{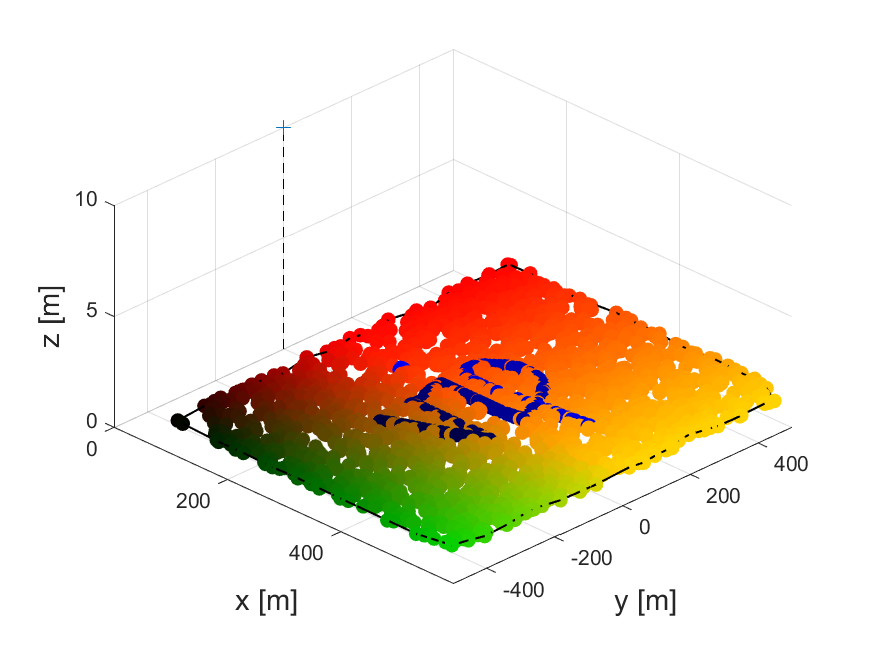} &
  \end{tabular}
  \caption{3D environment.}
  \label{fig31}
\end{figure*}
\else
\begin{figure}[!t]
\vspace{-7.5mm}
\centering
  \renewcommand{\arraystretch}{0}%
  \begin{tabular}{@{}c@{\hspace{1pt}}c@{\hspace{1pt}}c@{}}
  \includegraphics[width=3.0in]{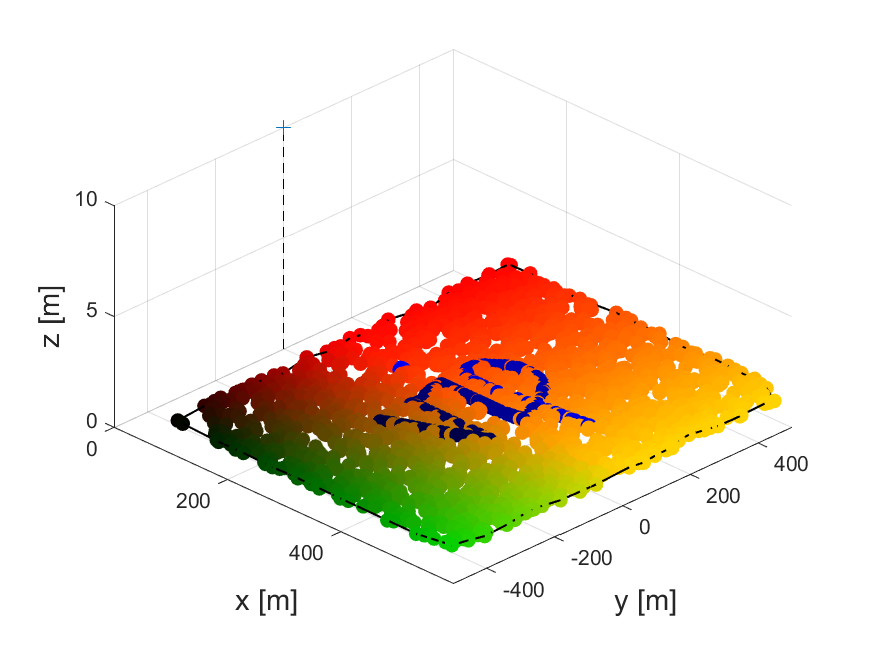} &
  \end{tabular}
  \caption{3D environment.}
  \label{fig31}
\end{figure}
\fi
We employed the simulation environment in \cite{b1} with the purpose of comparing the performance in a fair fashion, as in \cite{AA24}. The simulation parameters we used are given in Table~\ref{tab1} at SNR = 0 dB. SNR is defined as the ratio of the signal power to the noise power and 0 dB means that the two are equal.
%In the first scenario, we assume the antenna elements and the UEs are in the same plane as in Figure~\ref{fig30}, and in the second scenario
We used a three-dimensional environment exactly as in paper \cite{b1} as shown in Fig.~\ref{fig31}, where the antenna is 8.5 meters above the plane of the UEs. %We call the first scenario 2D, and the second scenario 3D.
%We call this three-dimensional scenario as 3D. A similar two-dimensional (2D) scenario is investigated in \cite{Aly22}.
The simulation environment is 1000m $\times$ 500m. The 2048 UEs are placed randomly, except 234 of the UEs are selected to make the word ``VIP," so we can see if the channel chart preserves the shape.

%% file: performance.tex
%
%\section{Basis for Performance Comparison}\label{ch:4}
%
%\subsection{Continuity and Trustworthiness}\label{sec:CT+TW}
%
In addition to a visual comparison of channel charts, we will use continuity (CT) and trustworthiness (TW) as objective performance measures \cite{b1,AA24}. CT specifies if neighbors in the original space are close in the representation space.  TW measures how well the feature mapping avoids introducing new neighbor relations that were not present in the original space. 
\begin{comment}
Let ${\cal V}_K({\bf u}_i)$ be the $K$-neighborhood of point ${\bf u}_i$ in the original space. 
Also, let $\hat r(i,j)$ be the ranking of point ${\bf v}_j$ among the neighbors of point ${\bf v}_i$, ranked according to their similarity to ${\bf v}_i$. Then the point-wise continuity of the representation ${\bf v}_i$ of the point ${\bf u}_i$ is defined as
\begin{equation}
{\rm CT}_i(K) = 1 - \frac{2}{K(2N-3K -1)} \sum_{j\in {\cal V}_K({\bf u}_i)}({\hat r}(i,j) -K).
\end{equation}
The (global) continuity of a point set $\{{\bf u}_n\}_{n=1}^N$ and its representation $\{{\bf v}_n\}_{n=1}^N$ is
\begin{equation}
{\rm CT}(K)=\frac{1}{N}\sum_{i=1}^N{\rm CT}_i(K).
\end{equation}
Now, let ${\cal U}_K({\bf v}_i)$ be the set of ``false neighbors'' that are in the $K$-neighborhood of ${\bf v}_i$, but not of ${\bf u}_i$ in the original space. Also, let $r(i,j)$ be the ranking of point ${\bf u}_i$ in the neighborhood of point ${\bf u}_i$, ranked according to their similarity to ${\bf u}_i$. The point-wise trustworthiness of the representation of point ${\bf u}_i$ is then
\begin{equation}
{\rm TW}_i(K)=1-\frac{2}{K(2N-3K-1)}\sum_{j\in {\cal U}_K({\bf v}_i)}(r(i,j)-K).
\end{equation}
The (global) trustworthiness between a point set $\{{\bf u}_n\}_{n=1}^N$ and its representation $\{{\bf v}_n\}_{n=1}^N$ is
\begin{equation}
{\rm TW}(K)=\frac{1}{N}\sum_{i=1}^N{\rm TW}_i(K).
\end{equation}
\end{comment}
For mathematical descriptions of point-wise and global CT and TW, we refer the reader to \cite{b1,AA24}
Point-wise and global CT and TW are between 0 and 1, with larger values being better \cite{b1}.

%% file: AoAAlgorithms.tex
\section{Algorithms for AoA Estimation}
We will discuss three relatively simple algorithms on AoA estimation from the literature. For an introduction to this subject, see, e.g., \cite{VanTrees,KV96}.
\subsection{Bartlett's Algorithm}
With our earlier definition of the autocorrelation matrix ${\bf R}$ and the steering vector $A(\theta)$, this algorithm computes
\begin{equation}
P_{\it Bartlett}(\theta) = \frac{{\bf A}^H(\theta) {\bf R}{\bf A}(\theta)}{{\bf A}^H(\theta){\bf A}(\theta)}
\end{equation}
and then finds the maximum of $P_{\it Bartlett}(\theta)$. For a ULA, the denominator is a constant, and it is sufficient to work with
\begin{equation}
P_{\it Bartlett}(\theta) = {\bf A}^H(\theta) {\bf R}{\bf A}(\theta).
\end{equation}
For a derivation of this algorithm via optimization, see, e.g., \cite{KV96}.
\subsection{MVDR (Capon's) Algorithm}
The full name for this algorithm is Minimum Variance Distortionless Algorithm (MVDR). It is also know as Capon's algorithm. Its spatial spectrum $P_{\it MVDR}(\theta)$, similar to $P_{\it Bartlett}(\theta)$, is given as
\begin{equation}
P_{\it MVDR}(\theta)=\frac{1}{{\bf A}^H(\theta){\bf R}^{-1}{\bf A}(\theta)}.
\end{equation}
Once again, after calculation, one searches for the maximum of $P_{\it MVDR}(\theta)$ to determine the AoA.
\subsection{Minimum Norm Algorithm}
This algorithm carries out an eigenspace analysis as in MUSIC. It performs the eigenvalue decomposition
\begin{equation}
{\bf R} = {\bf U}\boldsymbol{\Sigma}{\bf U}^H = \left[ {\bf U}_s\ \; {\bf U}_n \right]
\left[\begin{array}{cc}{\bf D}_s&{\bf 0}\\ {\bf 0}&\sigma^2{\bf I}\end{array}\right]
\left[{\bf U}_s\ \; {\bf U}_n \right]^H
\end{equation}
where ${\bf U}_s$ corresponds to the signal subspace, consisting of $K$ signal eigenvectors; ${\bf U}_n$ corresponds to the noise subspace, consisting of $N-K$ noise eigenvectors; ${\bf D}_s$ is a $K\times K$ diagonal matrix with signal eigenvalues as the entries; and ${\bf I}$ is an $(N-K)\times (N-K)$ identity matrix. Then, the Minimum Norm Algorithm computes the spatial spectrum as
\begin{equation}
P_{\it MinNorm}(\theta)=\frac{1}{{\bf A}^H(\theta){\bf U}_n{\bf U}_n^H{\bf e}{\bf e}^H{\bf U}_n{\bf U}_n^H{\bf A}(\theta)}
\end{equation}
where ${\bf e}$ is the first column of the identity matrix of size $N\times N$. 

%% file: simulationresults.tex
\section{Simulation Results}
\begin{figure}[!t]
\centering
\includegraphics[width=0.5\textwidth]{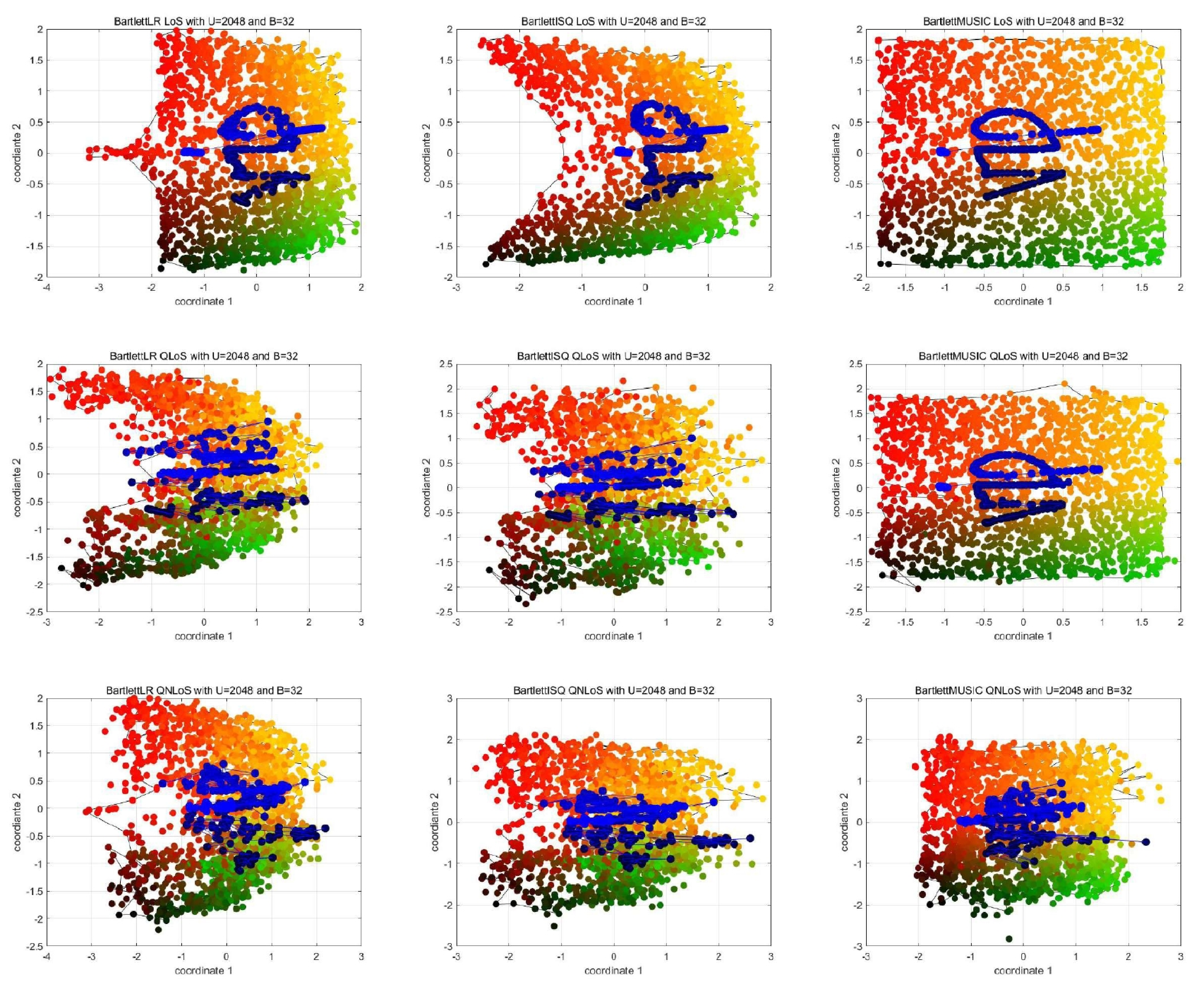}
\caption{Channel charts with Bartlett algorithm for $\theta$ and LR (left column), ISQ (middle column), and MUSIC (right column) algorithms for $\rho$ for the 3D LOS (top row), QLOS (middle row), and QNLOS (bottom row) channels.}
\label{fig:BLIM}
\end{figure}

\begin{figure}[!t]
\centering
\includegraphics[width=0.5\textwidth]{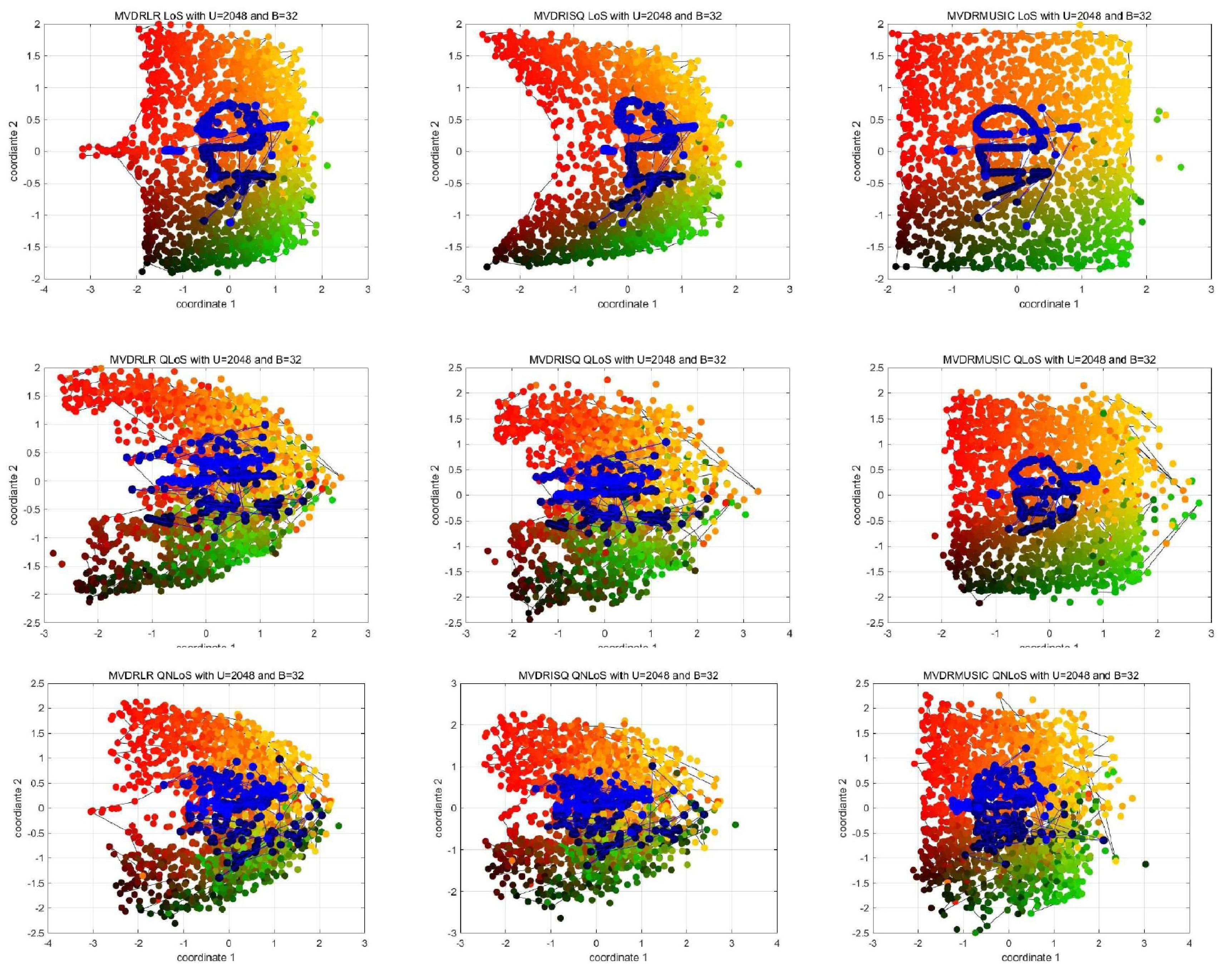}
\caption{Channel charts with MVDR algorithm for $\theta$ and LR (left column), ISQ (middle column), and MUSIC (right column) algorithms for $\rho$ for the 3D LOS (top row), QLOS (middle row), and QNLOS (bottom row) channels.}
\label{fig:MVDRLIM}
\end{figure}

\begin{figure}[!t]
\vspace{1mm}
\centering
\includegraphics[width=0.5\textwidth]{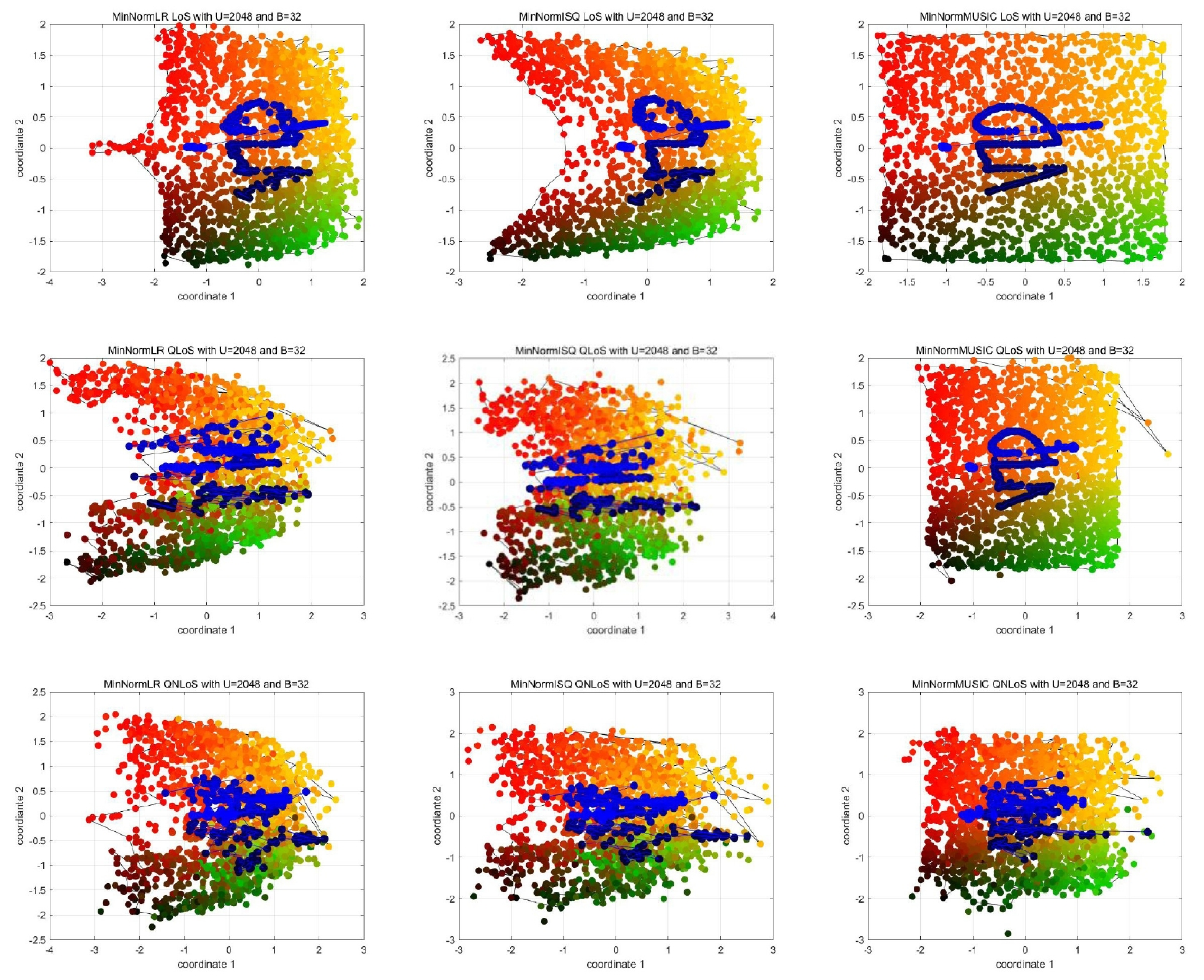}
\caption{Channel charts with Minimum Norm algorithm for $\theta$ and LR (left column), ISQ (middle column), and MUSIC (right column) algorithms for $\rho$ for the 3D LOS (top row), QLOS (middle row), and QNLOS (bottom row) channels.}
\label{fig:MNormLIM}
\end{figure}

We will first consider the ability of the Bartlett, MVDR, and Minimum Norm AoA algorithms to estimate $\theta$ and one of LR, ISQ, and MUSIC algorithms to estimate $\rho$. The results in terms of channel charts are given in Figs~\ref{fig:BLIM}-\ref{fig:MNormLIM}. Based on these charts, one can state that for any of Bartlett, MVDR, and Minimum Norm algorithms for estimating $\theta$, the MUSIC algorithm to estimate $\theta$ appears to be the best, while the LR and ISQ algorithms not appearing performing much different from each other.


On the other hand, Table~\ref{tbl:tab02} tabulates the execution runtimes of the algorithms in Fig.~\ref{fig:BLIM}--\ref{fig:MNormLIM}. According to this table, the execution times of MUSIC for $\theta$ are much larger than those of LR and ISQ. It can be seen that in execution times, using MUSIC for estimating $\rho$ causes a large runtime, while using LR or ISQ have similar runtimes. On the other hand, among Bartlett, MVDR, and Minimum Norm for estimating $\theta$, Minimum Norm has a very large runtime against Bartlett and MVDR. This means, we can eliminate Minimum Norm from consideration in estimating $\theta$.
\begin{table}[!h]
\caption{Execution runtimes of the algorithms in Figs.~\ref{fig:BLIM}--\ref{fig:MNormLIM}. The units are in seconds. The results are the averages of three runs.}
\label{tbl:tab02}
\begin{tabular}{|l|c|c|c|}
\hline
&Bartlett/LR&Bartlett/ISQ&Bartlett/MUSIC\\
\hline
LoS&0.4212&0.4154&16.0008\\
QLoS&0.4330&0.4113&15.9934\\
QNLoS&0.4789&0.4679&15.8993\\
\hline
&MVDR/LR&MVDR/ISQ&MVDR/MUSIC\\
\hline
LoS&6.4927&6.3431&21.0445\\
QLoS&6.4952&6.4381&21.2382\\
QNLoS&6.4735&6.4374&21.2727\\
\hline
&Min. Norm/LR&Min. Norm/ISQ&Min. Norm/MUSIC\\
\hline
LoS&13.5298&13.1709&28.1667\\
QLoS&12.8640&13.1466&27.7734\\
QNLoS&13.0934&13.0816&28.0243\\
\hline
\end{tabular}
\end{table}

\begin{figure}[!h]
\centering
\includegraphics[width=0.5\textwidth]{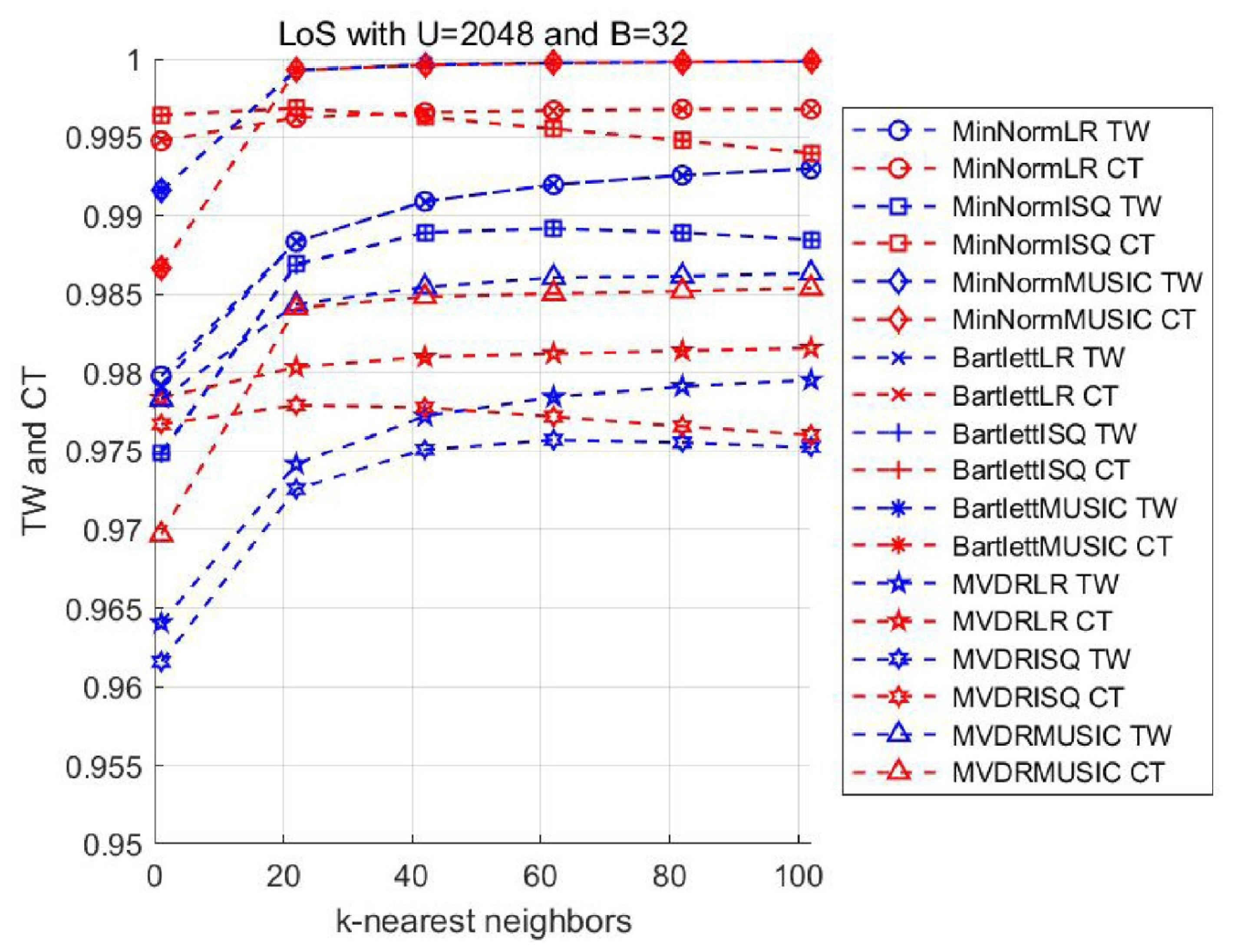}
\caption{TW and CT values for an LOS channel.}
\label{fig:TWCT-LOS}
\end{figure}
\begin{figure}[!h]
\centering
\includegraphics[width=0.5\textwidth]{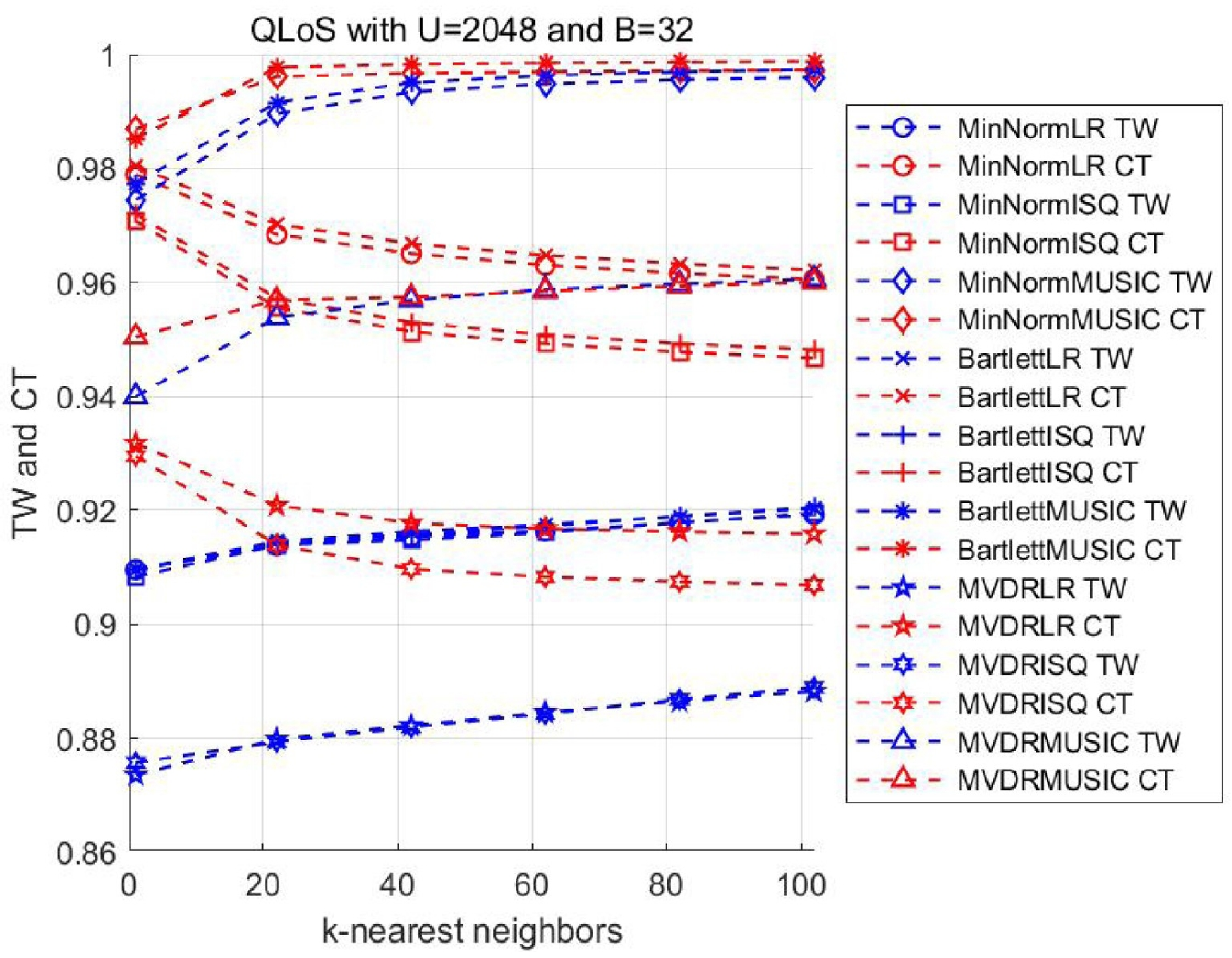}
\caption{TW and CT values for a QLOS channel.}
\label{fig:TWCT-QLOS}
\end{figure}
\begin{figure}[!h]
\centering
\includegraphics[width=0.5\textwidth]{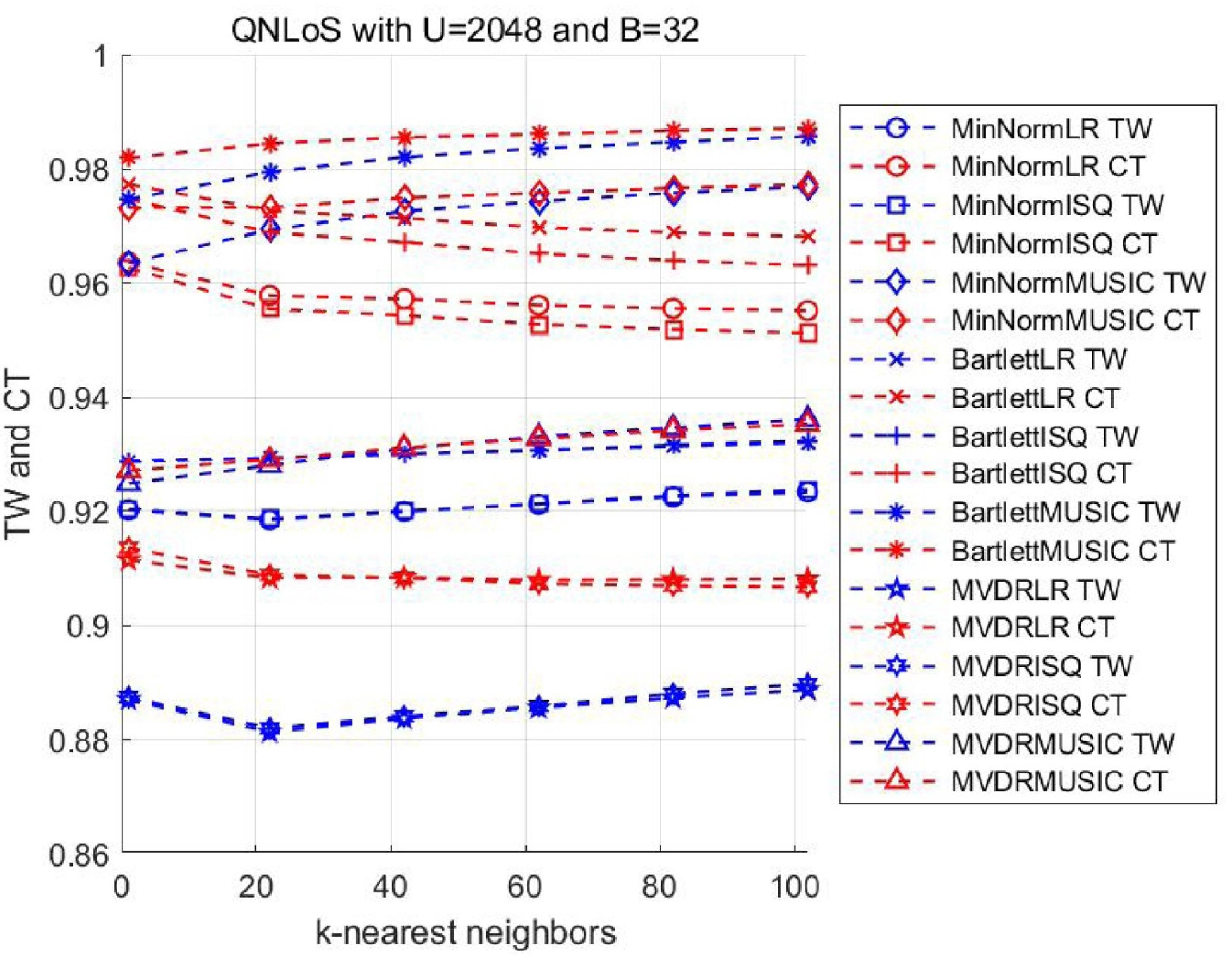}
\caption{TW and CT values for a QNLOS channel.}
\label{fig:TWCT-QNLOS}
\end{figure}
We tabulate the TW and CT values of the algorithms in Figs.~\ref{fig:BLIM}--\ref{fig:MNormLIM} for LOS, QLOS, and QNLOS channels at 102 nearest points in Table~\ref{tbl:tab03}. Since we have already eliminated Minimum Norm algorithm, we conclude from this table that the Bartlett algorithm may have an edge over the MVDR algorithm in terms of TW and CT performance. To investigate this further, we consider Figs.~\ref{fig:TWCT-LOS}--\ref{fig:TWCT-QNLOS} in LOS, QLOS, and QNLOS channels for $k$ nearest neighbors for values of $k$ in the range 0--102. A careful study of these plots indicate that, in terms of TW and CT performance, Bartlett algorithm can indeed be a contender even though it can be implemented in a simple fashion.
\begin{table*}[!t]
\centering
\small
\caption{TW and CT values of the algorithms in Figs.~\ref{fig:BLIM}--\ref{fig:MVDRLIM} for LOS, QLOS, and QNLOS channels.}
\label{tbl:tab03}
\begin{tabular}{|c|l|c|c|c|c|c|c|c|c|c|}
\hline
Measure&Channel&Bartlett&Bartlett&Bartlett&MVDR/&MVDR/&MVDR/&MinNorm/&MinNorm/&MinNorm/\\
&&LR&ISQ&MUSIC&LR&ISQ&MUSIC&LR&ISQ&MUSIC\\
\hline
\multirow{3}{*}{TW}&LOS&0.9930&0.9885&0.9998&0.9796&0.9753&0.9864&0.9330&0.9885&0.9998\\
&QLOS&0.9203&0.9205&0.9975&0.8882&0.8889&0.9607&0.9192&0.9194&0.9960\\
&QNLOS&0.9322&0.9324&0.9857&0.8887&0.8898&0.9362&0.9234&0.9238&0.9769\\
\hline
\multirow{3}{*}{CT}&LOS&0.9968&0.9940&0.9998&0.9816&0.9761&0.9864&0.9968&0.9940&0.9998\\
&QLOS&0.9622&0.9483&0.9989&0.9158&0.9068&0.9601&0.9606&0.9468&0.9974\\
&QNLOS&0.9682&0.9631&0.9872&0.9082&0.9067&0.9354&0.9552&0.9513&0.9773\\
\hline
\end{tabular}
\end{table*}

\begin{figure}[!h]
\centering
\includegraphics[width=0.5\textwidth]{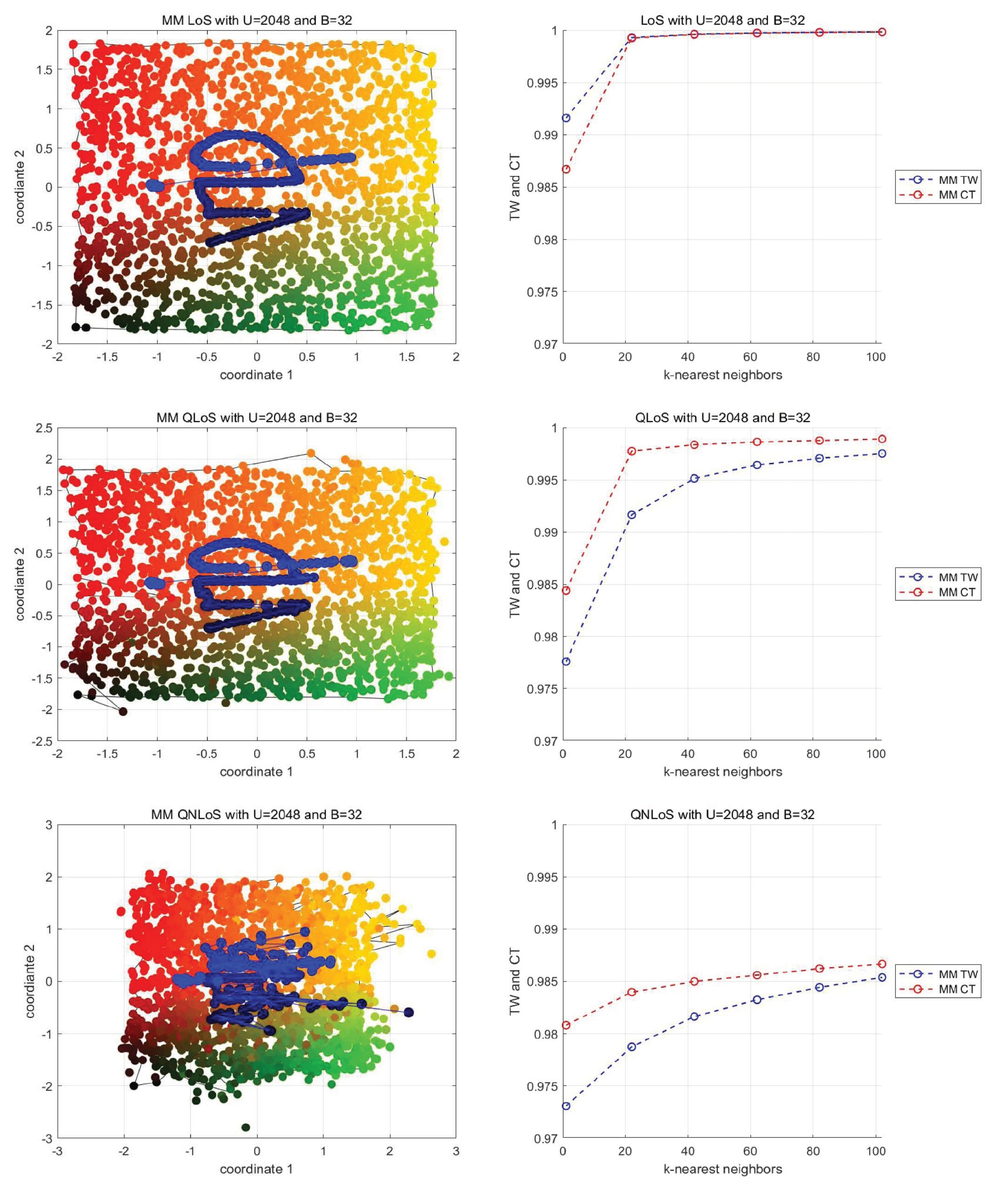}
\caption{Channel charts and TW and CT values for employing the MUSIC algorithm to estimate both $\theta$ and $\rho$.}
\label{fig:MM}
\end{figure}

\begin{figure}[!h]
\centering
\includegraphics[width=0.494\textwidth]{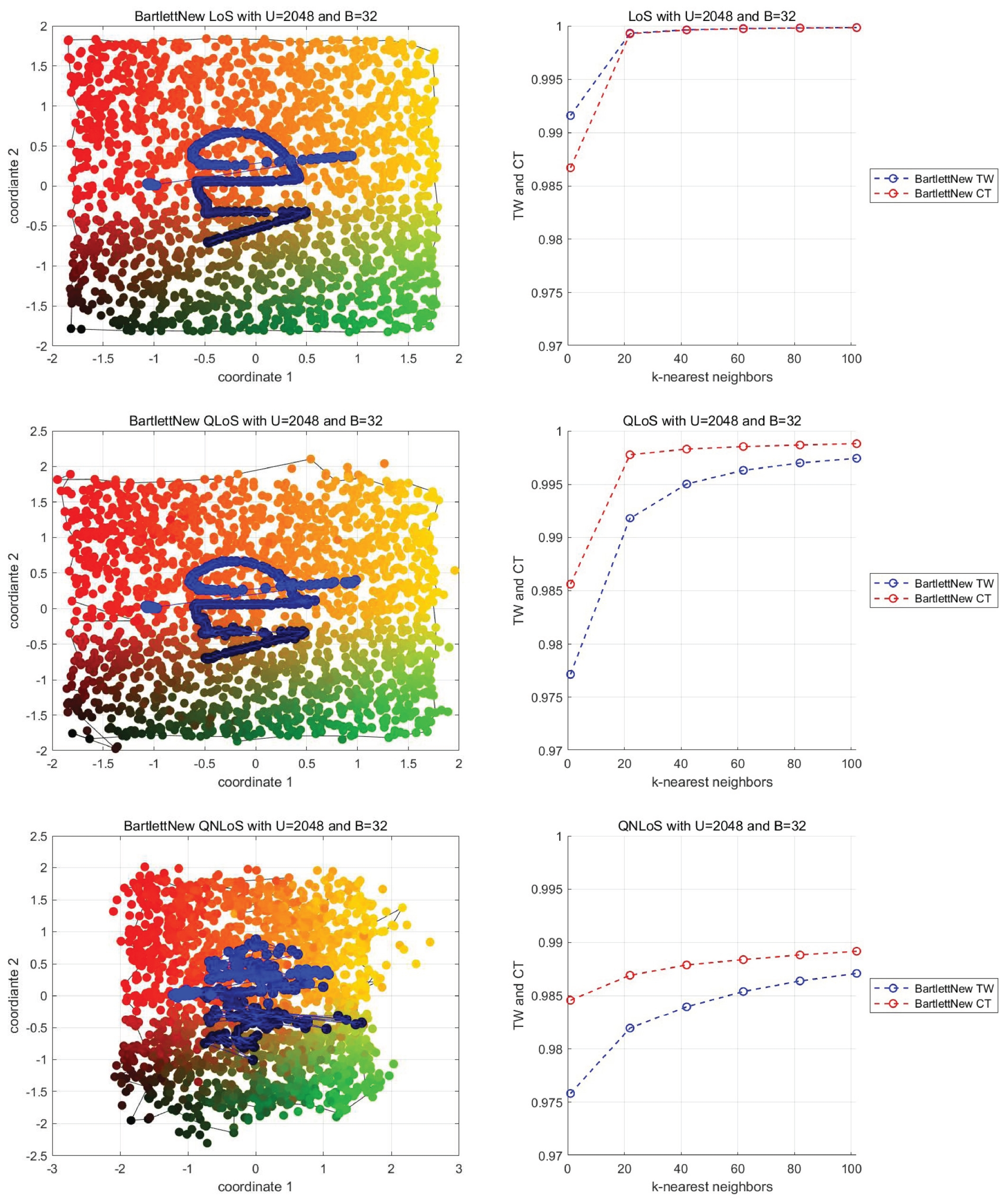}
\caption{Channel charts and TW and CT values for employing the Bartlett algorithm to estimate both $\theta$ and $\rho$.}
\label{fig:BB}
\end{figure}

In our work \cite{AA24}, we came to the conclusion that using the MUSIC algorithm for estimating both $\theta$ and $\rho$ among the algorithms studied in that paper, including LR and ISQ. We call the resulting algorithm MUSIC/MUSIC, or MM. We show the performance of employing the MM algorithm in Fig.~\ref{fig:MM}. In Table~\ref{tbl:tab04} we provide TW and CT values at 102 nearest points using the MM algorithm. We provide the running times of using the MM algorithm in Table~\ref{tbl:tab05}.
\begin{table}
\centering
\caption{TW and CT values at 102 nearest points for employing the MUSIC algorithm to estimate both $\theta$ and $\rho$.}\label{tbl:tab04}
\begin{tabular}{|l|l|c|}
\hline
Measure&Channel&MUSIC/MUSIC\\
\hline
\multirow{3}{*}{TW}&LOS&0.9998\\
&QLOS&0.9975\\
&QNLOS&0.9854\\
\hline
\multirow{3}{*}{CT}&LOS&0.9998\\
&QLOS&0.9989\\
&QNLOS&0.9867\\
\hline
\end{tabular}
%\end{table}
%
%\begin{table}
\vspace{3mm}
\centering
\caption{Running times for employing the MUSIC algorithm to estimate both $\theta$ and $\rho$.}\label{tbl:tab05}
\begin{tabular}{|l|c|}
\hline
Channel&Runtime (s)\\
\hline
LOS&20.6577\\
QLOS&20.6061\\
QNLOS&20.8503\\
\hline
\end{tabular}
\end{table}
We show the performance of employing the Bartlett algorithm for estimation of both $\theta$ and $\rho$ in Fig.~\ref{fig:BB}. In Table~\ref{tbl:tab06} we provide TW and CT values at 102 nearest points using Bartlett algorithm for both $\theta$ and $\rho$. We provide the running times of using the Bartlett algorithm to estimate both $\theta$ and $\rho$ in Table~\ref{tbl:tab07}. In implementing the Bartlett algorithm, Cholesky factorization of the autocorrelation matrix ${\bf R}$ is made for reduction in computational complexity. From the results presented, clearly the Bartlett algorithm has a substantial reduction in computational complexity without a discernible reduction in the quality of the channel charts or performance measures TW and CT.
\begin{table}
\centering
\caption{TW and CT values at 102 nearest points for employing the Bartlett algorithm to estimate both $\theta$ and $\rho$.}\label{tbl:tab06}
\begin{tabular}{|l|l|c|}
\hline
Measure&Channel&Bartlett/Bartlett\\
\hline
\multirow{3}{*}{TW}&LOS&0.9998\\
&QLOS&0.9974\\
&QNLOS&0.9871\\
\hline
\multirow{3}{*}{CT}&LOS&0.9998\\
&QLOS&0.9988\\
&QNLOS&0.9892\\
\hline
\end{tabular}
%\end{table}
%
%\begin{table}
\vspace{3mm}
\centering
\caption{Running times for employing the Bartlett algorithm to estimate both $\theta$ and $\rho$.}\label{tbl:tab07}
\begin{tabular}{|l|c|}
\hline
Channel&Runtime (s)\\
\hline
LOS&1.1203\\
QLOS&1.1407\\
QNLOS&1.1373\\
\hline
\end{tabular}
\end{table} 

%% file: conclusion.tex
\section{Conclusion}\label{ch:5}
The LR, ISQ, and MM algorithms we presented in \cite{AA24} significantly outperformed the three algorithms in the seminal paper \cite{b1}, PCA, SM, and AE, in terms of performance. In this paper, we investigated the performance of the more conventional AoA estimation algorithms Bartlett, Minimum Variance Distortion Response (MVDR or Capon), and Minimum Norm algorithms. As in \cite{b1}, we measured the performance in terms of the visual appearance of the channel charts, as well as connectivity (CT) and trustworthiness (TW). We also considered execution time or running time of the algorithms. In our study of the use of the MUSIC algorithm to estimate $\theta$ and one of Bartlett, MVDR, or Minimum Norm algorithms, we conclude that all three conventional algorithms can be competitive.